\documentclass[runningheads]{llncs}
\usepackage[T1]{fontenc}
\usepackage{arydshln}

\usepackage{algorithm2e}
\SetAlFnt{\scriptsize}
\usepackage{amsmath}
\usepackage{xcolor}

\usepackage{graphicx}
\begin{document}
\title{Relational Database Data Lineage Ontology}
\author{Jakub Dutkiewicz\inst{1}\orcidID{0000-0002-7954-7484} \and
Pawel Misiorek\inst{1}\orcidID{0000-0001-5223-240X} \and
Robert Wrembel\inst{1}\orcidID{0000-0001-6037-5718}}
\authorrunning{J. Dutkiewicz et al.}
\institute{Poznan University of Technology, Poznan, Poland \\
\email{\{firstname.lastname\}@put.poznan.pl}}
\maketitle              % typeset the header of the contribution
\begin{abstract}
Modeling data lineage in relational databases remains a challenging problem, particularly in scenarios involving incomplete or missing dependencies between database objects. In this paper, we propose a novel ontology for relational database data lineage, designed to provide a richer and more expressive semantic representation supporting discovering the lineage links by means of knowledge graphs (KGs). Building upon our previous work on KG-based lineage discovery, the proposed ontology extends the earlier model with additional concepts capturing structural, semantic, and transformation-level characteristics of relational data. These extensions enable more precise encoding of lineage evidence. To evaluate the impact of the proposed ontology, we conduct a comparative study using a KG-based inductive link prediction framework. Specifically, we assess the performance of a graph neural network model based on path embeddings under two settings: using the original baseline ontology and the newly proposed one. 
Experimental results demonstrate that the application of the enriched semantic model leads to improvements in lineage link prediction performance, as measured by AUC and Hits@10 metrics.
% , while the attention-based model further enhances predictive accuracy. 
%These findings highlight the importance of ontology design in KG-driven data lineage discovery.

\end{abstract}

\section{Introduction}
%========================================================
Data lineage refers to the set of techniques that document the life-cycle of data, from their origin, through transformations, to their final form in a target system~\cite{IBMLin}. In some business domains, like the financial sector, regulatory acts require a full documentation on how key data were processed and used, e.g., to assess credit risk or in reports to supervisory bodies.

A particular instance of data lineage is \textit{object lineage}, which captures dependencies between database (DB) objects such as tables, views, materialized views, and other specific objects. These dependencies are crucial for ensuring data consistency, supporting debugging and auditing processes, as well as sharing data and query results between various applications. However, in practice, many lineage links (relationships) remain incomplete or entirely missing due to complex transformations, temporary objects, or undocumented data processing pipelines. Such missing lineage links will further be called as \textit{broken lineage}.

A natural way to represent lineage relationships is through graph-based models, where nodes correspond to DB objects and edges encode dependencies between the objects. In recent years, Knowledge Graphs (KGs) have been widely used as a useful framework for representing structured and semi-structured data with rich semantics~\cite{hogan2021}. In parallel, significant advances have been made in inductive link prediction methods, particularly those based on Graph Neural Networks (GNNs), which enable the discovery of missing relationships in graphs~\cite{tdh20,mu25,wfw24}. Despite this progress, the role of semantic modeling, and in particular ontology design, remains underexplored in the context of data lineage discovery.

In our previous work~\cite{dutkiewicz2026}, we introduced a novel framework for discovering missing data lineage links by transforming relational databases into KGs using an ontology-grounded semantic model. We proposed a benchmark dataset of lineage scenarios, a transformation procedure from relational schemas to KGs, and an inductive link prediction approach based on a path-based GNN model. Our experiments demonstrated that such a transformation enables effective discovery of broken lineage links, highlighting the potential of combining semantic technologies with machine learning for this task.

In this paper, we extend this research by focusing on the ontology itself as a key component of a lineage discovery pipeline. We propose a new, extended ontology for discovering broken data lineage links in relational DBs, designed to capture structural and semantic characteristics of relational data and transformations. It has to be stressed that there is limited work on designing ontology models specifically tailored to enable machine learning-based lineage inference in relational databases. In particular, the impact of ontology design on the quality of algorithms addressing tasks such as link prediction remains largely unexplored. Our work addresses this gap by proposing an extended ontology that combines relational data modeling with provenance semantics.

The main contributions of this work are as follows: (1) we introduce an enhanced ontology model for representing data lineage in relational DBs, extending our previous semantic framework \cite{dutkiewicz2026} with additional concepts and relationships, (2) we provide a comparative evaluation of the impact of ontology design on lineage link prediction, by applying the the GNN-based model under both the original and the proposed ontology.
% , and (3) we investigate the effectiveness of a more advanced Graph Attention Network (GAT)-based model in combination with our extended ontology. The experimental results demonstrate that both the enriched ontology and the attention-based architecture contribute to improved performance in broken lineage discovery.

\section{Related Work}
%========================================================
The contribution of this paper merges three research tracks, namely: (1) discovering data lineage, (2) link prediction in KGs, and (3) modeling ontologies. All these three tracks are briefly outlined in this section.

\subsection{Data Lineage/Provenance}
%-----------------------------------
A few types of solutions for storing/retrieving data provenance in DBs have been proposed, namely: annotation-based, inversion-based, and lineage graphs. In the \textit{annotation-based} solutions input data are annotated with information about their provenance (for example a location in which they are stored) and the annotations are propagated throughout the whole data processing workflow, until the final destination. Historically, first annotation-based solutions include Polygen \cite{WM90} and DBNotes \cite{Chi05}, whereas more recent works include \cite{FGT08,DDS20}. A common approach to annotation based data provenance representation is the \textit{semiring} model \cite{GKT07,GT17}. The \textit{inversion-based} solutions invert queries trying to infer original data that produced a given result, e.g., \cite{CWW00,YKAM23}. Finally, in the \textit{lineage graph} approach, provenance is registered during query processing and saved in the form of a lineage graph for further analysis \cite{Kas20}. Paper \cite{Sen18} summarizes several most relevant works on provenance techniques. 

Another path includes solutions for tracking provenance in scripting languages (e.g., Python and R), e.g., \cite{SGB14,PFMB19,CMST20}. \cite{PMBF17} proposes automatic Python processing without modifying existing scripts. Similar assumptions were made in \cite{NFP+20}, dedicated for data science applications. A completely different approach was proposed by the authors of~\cite{LC22b}, who advocate for allowing humans to express the desired provenance through a provenance schema. 

%In order to support the exchange of provenance data between different systems, the Open Provenance Model (OPM) was created \cite{MFFMMP2008}. The W3C standard PROV \cite{MBC2013} aims at facilitating provenance data exchange over the Web. 

To the best of our knowledge, discovering missing provenance links in relational DBs has been explicitly addressed only in \cite{BR25}. To this end, the authors apply a neural network (NN) based on encoders. The solution is based on a simplified assumption that column names between related objects are similar.

\subsection{Inductive Link Prediction for Knowledge Graphs}
%-----------------------------------
The surveys in \cite{dna24-survey,rdd21-survey-kg} provide comprehensive overviews of link prediction methods for graphs and knowledge graphs (KGs), respectively. Traditional (transductive) approaches, such as TransE~\cite{b13-transe} and BERT-ConvE~\cite{liu22-bert-conve}, focus on inferring missing relations between entities observed during training. In contrast, inductive link prediction aims to generalize to previously unseen entities, enabling the prediction of new triples in a more practical scenario involving graphs with nodes unknown during the learning of the prediction model~\cite{mu25,tdh20}. Most KG link prediction methods rely on embedding techniques, where entities and relations are mapped into vector spaces~\cite{b13-transe,liu22-bert-conve}. While effective, these approaches typically struggle to generalize beyond the training graph. To address this limitation, more recent methods incorporate structural information to better capture semantic patterns~\cite{hyl2017}, though challenges remain in handling emerging entities. Recent advances in inductive link prediction tackle this issue by leveraging graph structure and relational patterns~\cite{mu25,tdh20,leskovec21,zl24-model}. Early approaches focused on rule-based reasoning~\cite{mel18}, while GraIL~\cite{tdh20} introduced a framework based on local subgraph representations for fully inductive settings. Subsequent work extended this idea by incorporating relational context and path-based features~\cite{leskovec21}, global relation graphs~\cite{mu25}, and neural architectures such as Siamese NN~\cite{zl24-model}. 

\subsection{Ontology Modeling for Data Lineage}
%-----------------------------------
Several standardized vocabularies have been proposed to model tabular and relational structures in a machine-readable form. In particular, the W3C CSV on the Web (CSVW) ontology~\cite{csvw} provides a framework for describing tabular data, including schema, columns, and data types, while the Data Set Description (DSD) vocabulary~\cite{dsd} enables the representation of structured datasets and their metadata. 
From the perspective of data lineage and provenance, the PROV-O ontology~\cite{Mor2015,provo} has become a widely adopted standard for modeling the origin, transformation, and dependencies of data \cite{chhetri25}. PROV-O introduces core concepts such as entities, activities, and agents, along with relations describing derivation and usage, making it a natural foundation for representing lineage information. Several works extend or adapt PROV-O to capture fine-grained provenance in databases and data processing pipelines~\cite{Pru2025}. However, these approaches typically focus on descriptive modeling and interoperability rather than supporting machine learning tasks. To the best of our knowledge, our research is the first to apply an ontology-grounded transformation of relational database to knowledge graph used for the task of missing lineage link discovery. 

\begin{figure}
\includegraphics[width=\textwidth]{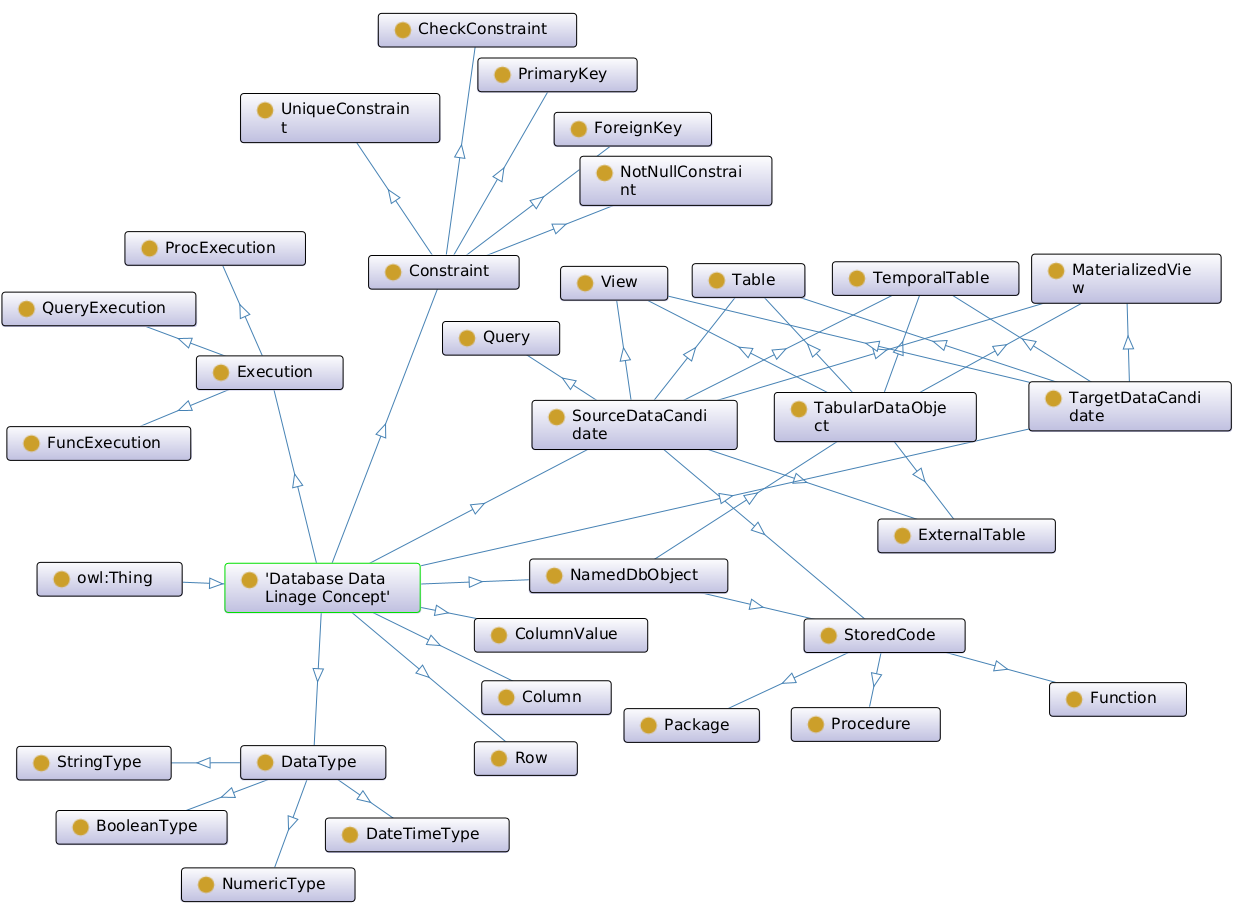}
\caption{The hierarchy of classes defined in the \textit{RDDL Ontology}; the hierarchy shows only proper classes of the \textit{RDDL Ontology}, i.e., it does not visualize the links to classes from standard ontologies (the graphic was generated using Prot\'eg\'e OntoGraf).} \label{fig:subclasses}
\end{figure}

\section{Relational Database Data Lineage Ontology}
%========================================================
In this work, we propose an extended ontology for modeling data lineage in relational databases, designed to provide a richer semantic representation than the model introduced in our previous study~\cite{dutkiewicz2026}. 

\begin{figure}
\includegraphics[width=0.99\textwidth]{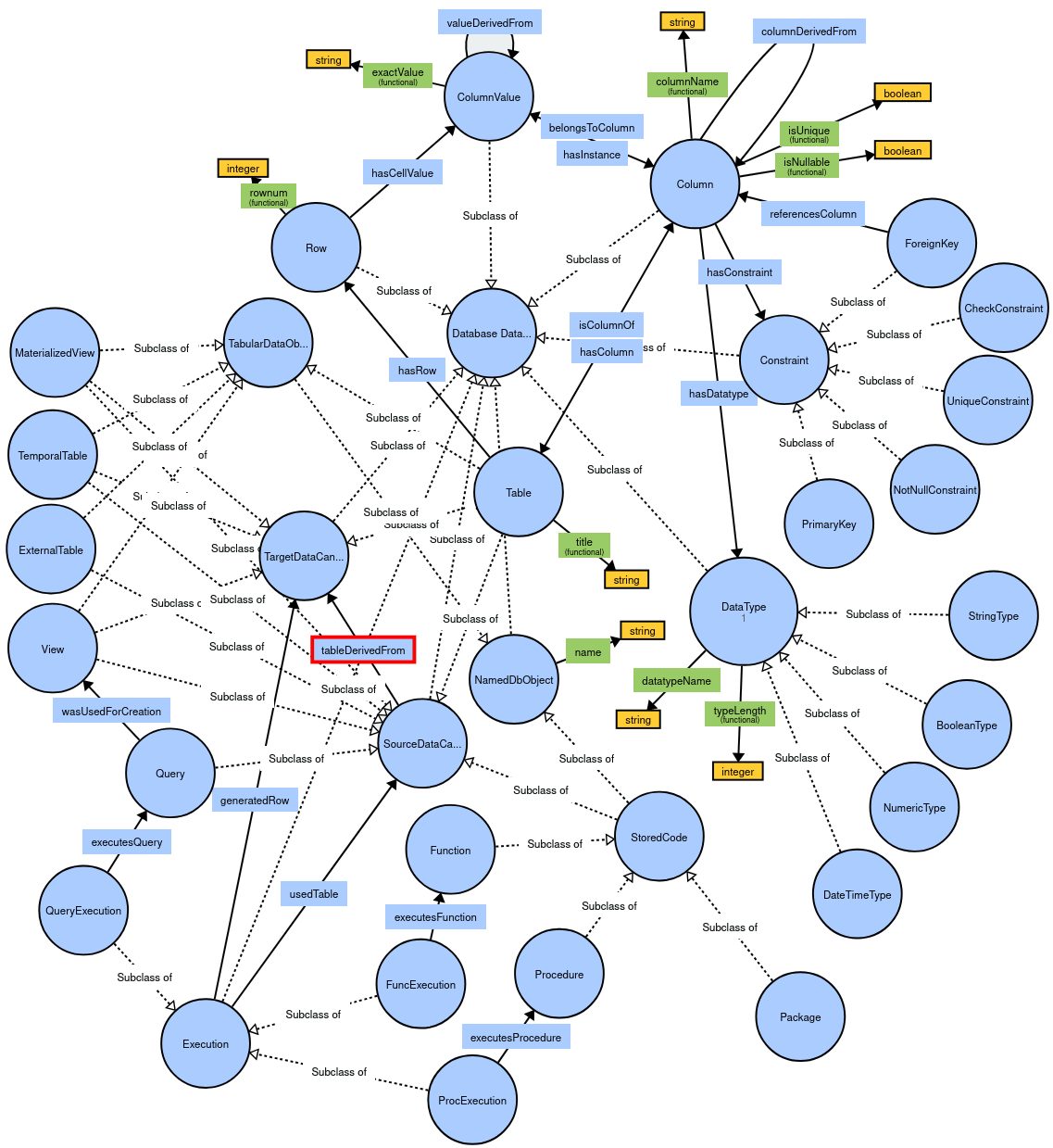}
\caption{The visualization of classes and relations of the \textit{RDDL Ontology} (the graphic was generated by WebVOWL).} 
\label{fig:relations}
\end{figure}

The goal of the Relational Database Data Lineage Ontology (\textit{RDDL Ontology}), which we propose, is to provide a semantic layer enabling the transformation of relational database schemas and selected data elements into graph structures that can be processed by graph NNs used for inductive link prediction. The ontology is defined in OWL and refers to concepts from several established vocabularies, including the CSV on the Web (CSVW) ontology \cite{csvw}, the Data Set Description (DSD) vocabulary \cite{dsd}, and the PROV-O provenance ontology \cite{provo}. These standards provide a semantic grounding for modeling tabular data structures and provenance relations, which are essential for representing data lineage dependencies between database objects. The \textit{RDDL Ontology} is available in the public repository\footnote{https://github.com/dudenzz/lineage/blob/main/ontology/rddl-ontology.ttl}. The visualizations of its classes hierarchy as well its object and data properties are presented on Fig.~\ref{fig:subclasses} and~\ref{fig:relations}, respectively.

\subsection{Core Concepts}
%-----------------------------------
The \textit{RDDL Ontology} models the essential components of relational databases using several core classes proposed in the baseline approach~\cite{dutkiewicz2026}. The central classes of the model include \textit{Table}, \textit{Column}, and \textit{Row}, and \textit{ColumnValue}, which represent structural components of relational data.
Each \textit{Table} is composed of multiple \textit{Column} instances that represent its attributes. 
%The relationship between tables and columns is represented using the \texttt{hasColumn} property.
\textit{Tables} may also contain multiple \textit{Row} instances representing records stored in a given table. To support fine-grained lineage modeling, the \textit{RDDL Ontology} also introduces class \textit{ColumnValue}, which represents individual data values stored in table rows. Each value is associated with a specific column using the \texttt{belongsToColumn} property and may participate in value-level lineage relations. 

In contrast to the baseline ontology proposed in~\cite{dutkiewicz2026}, the new ontology introduces an extended class hierarchy by adding new types of database objects such as \textit{View}, \textit{Query}, \textit{TemporalTable}, \textit{ExternalTable}, and \textit{MaterializedView}, as well as abstract classes including \textit{NamedDBObject}, \textit{StoredCode}, and \textit{TabularDataObject} (see Fig.~\ref{fig:subclasses}). This extension enables precise modeling of various types of database objects, which may be crucial for correctly identifying data lineage relationships.

\subsection{Lineage Modeling}
%-----------------------------------

The primary goal of our ontology is to support modeling of data lineage dependencies between DB objects. Lineage relationships are represented using properties derived from the PROV-O relation \texttt{prov:wasDerivedFrom}~\cite{provo}. This ensures compatibility with standard provenance semantics. To enable domain-specific reasoning, our model distinguishes multiple levels of lineage representation:

\begin{itemize}
\item \texttt{tableDerivedFrom} represents lineage relationships between tabular objects from class \textit{SourceDataCandidate} as a domain to class \textit{TargetDataCandidate} as a range (see Figures \ref{fig:subclasses} and \ref{fig:relations});
\item \texttt{columnDerivedFrom} represents derivation dependencies between columns;
\item \texttt{rowDerivedFrom} represents derivation dependencies between rows;
\item \texttt{valueDerivedFrom} represents lineage relations between individual data values being  instances of the \textit{ColumnValue} class.
\end{itemize}

This hierarchical modeling approach allows representing both coarse-grained lineage between DB tables and fine-grained dependencies between attributes and data values.

\subsection{Schema Constraints and Datatypes}
%-----------------------------------
The \textit{RDDL Ontology} also models structural metadata associated with relational schemas. Database constraints are represented using class \textit{Constraint} and its specializations, including \textit{PrimaryKey}, \textit{ForeignKey}, \textit{NotNullConstraint}, and \textit{CheckConstraint}. Constraints may be associated with tables or columns through the \texttt{hasConstraint} property. Foreign key relationships between tables are modeled using the \textit{ForeignKey} class and property \texttt{referencesTable}, allowing representation of structural dependencies between relational tables.

Column datatypes are modeled using class \textit{DataType} and its subclasses representing different datatype categories such as numeric, boolean, or temporal types. Datatypes are associated with columns via the \texttt{hasDatatype} property. 

\subsection{Modeling Data Transformation Processes}
%-----------------------------------
To capture procedural data transformations occurring within relational DBs, the proposed ontology introduces several classes representing execution processes. Class \textit{Query} refers to SQL queries whereas class \textit{StoredCode} represents database procedures, packages or functions responsible for generating or transforming data. Individual execution events are represented using classes \textit{QueryExecution}, \textit{ProcExecution} and \textit{FuncExecution}. These executions may reference input tables using the \texttt{usesTable} property and produce new output using the \texttt{generatesRow} relation.

Additionally, the \textit{RDDL Ontology} introduces relationships linking data elements to their structural context (e.g., \texttt{belongsToColumn}) and execution processes (e.g., \texttt{executesQuery}, \texttt{executesFunction}, \texttt{executesProcedure}). These properties provide explicit connections between data, schema, and transformation logic, which were not fully represented in the previous ontology.

These constructs enable representing operational aspects of database workflows and capturing provenance relationships between data elements generated during database operations.

\begin{table}[h]
\centering
\caption{Comparison of the baseline~\cite{dutkiewicz2026} and proposed ontology.} \label{tab:comp}
\begin{tabular}{p{3.3cm} p{4cm} p{4.5cm}}
\hline
\textbf{Aspect} & \textbf{Baseline ontology} & \textbf{RDDL Ontology} \\
\hline
Process representation & Implicit or absent & Explicit modeling of queries, functions, and procedures \\
\hline
Data types & Basic model & Dedicated classes, modeling names and length  \\
\hline
Execution semantics & Not represented & Dedicated classes for execution instances (e.g., QueryExecution) \\
\hline
DB object hierarchy & Not represented & Taxonomy grouping dedicated classes for different DB objects\\
\hline
Integrity constraints & Not represented & Dedicated classes (e.g., primary and foreign keys)  \\
\hline
\end{tabular}
\end{table}

\subsection{The Baseline Ontology vs. RDDL Ontology}
\label{simple}
%-----------------------------------
The \textit{RDDL Ontology} (proposed in this paper) introduces the following important extensions that increase both expressiveness and applicability for machine learning tasks, as compared to the baseline ontology proposed in \cite{dutkiewicz2026}. First, it adds the representation of integrity constraints, data types, and hierarchical relationships between database objects. In particular, the ontology incorporates concepts for modeling attribute-level constraints such as primary keys, foreign keys, and uniqueness constraints, which play an important role in determining valid lineage paths and dependencies between data elements. Second, data types and their properties (name and length) are represented, allowing the model to capture compatibility and transformation semantics between source and target attributes. Third, the introduction of execution-related classes, such as \textit{QueryExecution}, \textit{FuncExecution}, and \textit{ProcExecution}, which represent the operational processes responsible for data transformations. These classes are linked to definitions of \textit{Query}, \textit{Function}, and \textit{Procedure}, respectively, allowing the ontology to capture not only structural dependencies but also the procedural context in which they arise. The differences between ontologies are summarized in Table~\ref{tab:comp}.

\begin{table}[h]
\caption{Main classes and object properties of the data lineage ontology.}
\label{tab:summary}
\centering
\small
\begin{tabular}{p{2.9cm} p{1.5cm} p{7.6cm}}
\hline
\textbf{Element} & \textbf{Type} & \textbf{Description} \\
\hline

\textit{Table} & Class & Represents a relational database table\\

\textit{Column} & Class & Represents a table attribute belonging to a table \\

\textit{Row} & Class & Represents a record stored in a relational table \\

\textit{CellValue} & Class & Represents an individual value stored in a table cell \\

\textit{DataType} & Class & Represents the datatype assigned to a column \\

\textit{Constraint} & Class & General class representing database constraints \\

\textit{PrimaryKey} & Class & Represents a primary key constraint \\

\textit{ForeignKey} & Class & Represents a foreign key constraint \\

\textit{StoredCode} & Class & Represents a stored procedure, package, or function \\

\textit{ProcExecution} & Class & Represents execution of a stored procedure \\

\textit{FuncExecution} & Class & Represents execution of a DB function \\

\hline

\texttt{hasColumn} & Property & Connects a table with its columns \\

\texttt{hasRow} & Property & Connects a table with its rows \\

\texttt{belongsToColumn} & Property & Connects a cell value to the column it belongs to \\

\texttt{hasDatatype} & Property & Associates a column with its datatype \\

\texttt{hasConstraint} & Property & Connects schema elements with defined constraints \\

\texttt{referencesTable} & Property & Indicates the referenced table in a foreign key constraint \\

\texttt{usesTable} & Property & Indicates that an execution uses a table as input \\

\texttt{generatesRow} & Property & Indicates that an execution generates a table row as output \\

\texttt{tableDerivedFrom} & Property & Represents lineage relationships between tables \\

\texttt{columnDerivedFrom} & Property & Represents lineage relationships between columns \\

\texttt{rowDerivedFrom} & Property & Represents lineage relationships between rows \\

\texttt{valueDerivedFrom} & Property & Represents lineage relations between cell values \\

\hline
\end{tabular}
\label{tab:ontology-summary}
\end{table}

Overall, the \textit{RDDL Ontology} provides a more expressive and semantically grounded representation of relational database data lineage. 
%By incorporating execution semantics and fine-grained dependencies, it enables improved feature representation for knowledge graph-based models, which is particularly beneficial for inductive link prediction tasks. 
The key classes and object properties of the proposed ontology are summarized in Table~\ref{tab:summary}.

%\subsection{Role of the RDDL Ontology in the Lineage Discovery Framework}
%-----------------------------------
%The ontology provides the semantic backbone of the proposed data lineage discovery framework. Relational databases used in training and testing experiments are transformed into KGs according to this ontology schema. The resulting KGs represent DB objects as nodes and structural or lineage relationships as edges. This unified representation enables the application of graph-based learning algorithms for identifying missing lineage relationships between database elements.

%RW - dotąd sprawdzone

\section{Experiments}
%========================================================
% \textcolor{red}{to trzeba przepisać bo nic z tego nie rozumiem} In order to validate the practical utility of the \textit{RDDL Ontology}, we designed a set of link prediction experiments.  We use a set of graphs embedded in the Ontology defined schemas. The graphs are created upon two variants of the Ontology - the \textit{Simplified Ontology} and \textit{RDDL Ontology}. An important aspect of the graph representation is an explicit expression of the \textit{data lineage} relationship. The graphs are prepared in an inductive manner - we use one graph for the preparation of the prediction model and a different graph for the validation of the model. Graphs are created upon various versions of the DB. The training and testing graphs share the same set of relationships, but nodes are completely distinct. 

To validate the practical utility of the \textit{RDDL Ontology}, we designed a series of data lineage discovery experiments. These tests utilize DB representation graphs built according to two ontology variants: the \textit{Baseline Ontology}~\cite{dutkiewicz2026} and the \textit{RDDL} \textit{Ontology}. The database objects in the graph, represented as nodes, are interconnected with \textit{data lineage} edges, representing transformations where new destination DB objects are derived from prior source DB objects. Following the inductive link prediction setting, each experiment utilizes two graphs with disjoint sets of nodes while sharing a common set of relationship types. The first graph explicitly contains the \textit{data lineage} relationships and is used for training the prediction model. In the second graph, used for testing, the \textit{data lineage} edges are hidden and serve only as ground truth for evaluation. 

The ontologies, implementation of the experiments, and the evaluation of the proposed method are available in the public repository\footnote{https://github.com/dudenzz/lineage}.

\subsection{Database Transformations}
%-----------------------------------
In order to simulate the missing data lineage, we provide a set of database transformation scenarios in the Northwind DB~\cite{northwind}, which include projections as well as linear and non-linear transformations such as bilinear, power, log and exponential transformations. On top of that, we propose three types of transformation scenarios:
\begin{itemize}
    \item selection-based scenarios - scenarios with only filtering conditions;
    \item join-based scenarios - scenarios with multiple, parallel data sources (e.g., \textit{Customers} and \textit{Order Details} tables used to create \textit{Customer Value} DB object);
    \item union-based scenarios - scenarios with multiple, equivalent data sources (e.g., \textit{Customers} and \textit{Employees} tables used to create \textit{People} DB object).
\end{itemize}

Each type of scenarios (e.g., selection-based non-linear transformation) is instantiated with 20 examples. Scenarios consist of four transformations, for which we track data lineage. The whole set of test scenarios includes 730 transformations. 

%\begin{table}[h]
%    \caption{Types of transformations (notation: $A'$ - a newly created attribute; $A, B$ - original attributes; $\Omega$ - the entire set of original attributes; $a,b,c$ - constant values).} 
%    \centering
%    \begin{tabular}{ccc}
%         \hline
%         Transformation & Classification & General formula  \\ \hline
%         Projection & Projection &$A'=\pi_A(R)$ \\
%         Linear transformation & Linear &$A' = \pi_{\sum_{A \subseteq  \Omega}A_i\cdot a_i}(R)$ \\
%         Bilinear transformation & Non-linear & $A' = \pi _{a_i \cdot AB + b}(R)$ \\
%         Power transformation & Non-linear & $A' = \pi _{a_i \cdot A^{c} + b}(R)$ \\
%         Log transformation & Non-linear & $A' = \pi _{a_i \cdot log_{c}A + b}(R)$ \\
%         Exponential transformation & Non-linear & $A' = \pi _{a_i \cdot c^{A} + b}(R)$ \\ \hline
%    \end{tabular}
%    \label{tab:transformationtypes}
%\end{table}

\subsection{Ontology-Based Knowledge Graph Construction}
%-----------------------------------
The conversion from the DB to the KG begins by querying the database for all schema objects, including table definitions and their respective integrity constraints. This structural metadata are initially stored in CSV files. To align with the graph-based data model, each $n$-ary relation is decomposed into a set of $n+1$ binary relations, facilitating the transition from tabular rows to ontological individuals. The logic governing this conversion is presented in Algorithm~\ref{alg1}.

\vspace{0.4cm}

\begin{algorithm}[H]
\DontPrintSemicolon
\SetKwFunction{FPopulate}{PopulateKnowledgeGraph}
\SetKwProg{Fn}{Function}{:}{}
\Fn{\FPopulate{$KG, Tables, Views, Config$}}{
    \ForEach{$V \in Views$}{
        $V_{ind} \gets KG.View(V.name)$\;
        \ForEach{$col \in V.columns$}{
            $C_{ind} \gets KG.Column(col.name)$\;
            $V_{ind}.hasColumn \cup \{C_{ind}\}$\;
            \If{$Config.use\_data$}{
                Map $V.rows$ to $Onto.Row$ and link $CellValue \to C_{ind}$\;
            }
        }
    }
    \ForEach{$T \in Tables$}{
        $T_{ind} \gets KG.Table(T.name)$\;
        \ForEach{$col \in T.columns$}{
            $C_{ind} \gets KG.Column(T.name + \text{"\_"} + col.name)$\;
            
                $C_{ind}.isNullable \gets col.nullable$\;
                $C_{ind}.hasDatatype \gets \text{ResolveType}(col.dtype, col.length)$\;

                \If{$col.isPK$}{ $C_{ind}.hasConstraint \cup \{KG.PrimaryKey(PK\_name)\}$ }
                \If{$col.isFK$}{ $C_{ind}.hasConstraint \cup \{KG.ForeignKey(FK\_name)\}$ }
            
        }
        
            \ForEach{$fk \in T.foreign\_keys$}{
                $target \gets KG.search(fk.ref\_table)$\;
                $KG.search(fk.name).referencesTable \cup \{target\}$\;
            }
        
    }
}
\caption{Converting a relational DB schema into a KG.}
\label{alg1}
\end{algorithm}

\vspace{0.4cm}

Once the low-level data are transformed into a KG, the system resolves the lineage tracked by SQL cursors during scenario execution. We represent a single entry of \textit{LineageData} as a tuple $[t_1, c_1, v_1, t_2, c_2, v_2]$, where $t_1$, $c_1$, and $v_1$ denote the lineage source table, column, and cell value, respectively, and $t_2$, $c_2$, and $v_2$ denote their corresponding counterparts in the lineage target. 
We use \textit{SPARQL} queries to identify the corresponding rows and inject the lineage information into the knowledge graph. This process is presented in Algorithm~\ref{alg2}.

\vspace{0.4cm}

\begin{algorithm}[H]
\DontPrintSemicolon
\SetKwFunction{FResolve}{ResolveLineage}
\SetKwProg{Fn}{Function}{:}{}
\Fn{\FResolve{$KG, LineageData$}}{
    \ForEach{$(t_1, c_1, v_1, t_2, c_2, v_2) \in LineageData$}{
        
        $row_{src} \gets \{ r \in KG : (r, \text{hasCellValue}, x) \land (x, \text{exactValue}, v_1) \land (x, \text{belongsToColumn}, c_1) \} \land (t_1, \text{hasColumn}, c_1) \}$\;
        
        $row_{dst} \gets \{ r \in KG : (r, \text{hasCellValue}, x) \land (cv, \text{exactValue}, v_2) \land (x, \text{belongsToColumn}, c_2)  \land (t_2, \text{hasColumn}, c_2) \}$\;

        $row_{dst}.\text{rowDerivedFrom} \cup \{row_{src}\}$\;

    }
}
\caption{Row-level lineage resolution via semantic triple matching.}
\label{alg2}
\end{algorithm}

% \begin{figure}
%     \centering
%     \includegraphics[width=0.6\textwidth]{figures/Structure.drawio.png}
%     \caption{Conversion graph base structure}
%     \label{fig:base_struct}
% \end{figure}

\subsection{Inductive Link Prediction Model}
%-----------------------------------
To perform link prediction in the KG, we implement a Multi-Path Siamese neural network\footnote{We used our own implementation that applies the most recent libraries.} based on the model proposed in~\cite{zl24-model}. This architecture is designed to encode complex topological relationships by processing multiple paths of edges simultaneously and comparing their aggregated representation against a target relationship embedding. By analyzing the connectivity patterns between nodes, the model effectively captures the structural features inherent in DB schemas.

\begin{figure}
    \centering
    \includegraphics[width=0.99\linewidth]{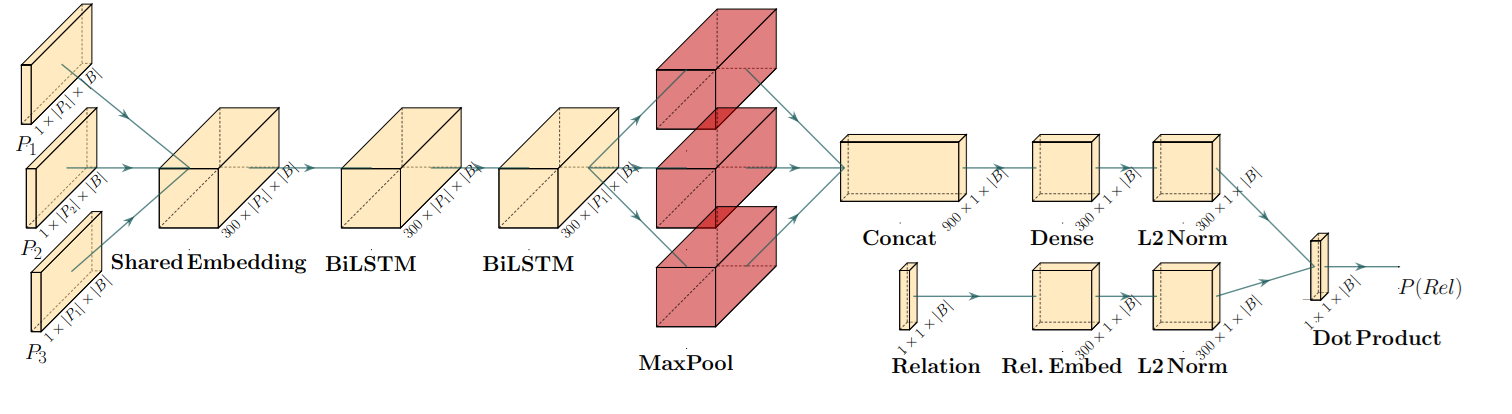}
    \caption{The neural network (NN) architecture. The model processes three paths ($P_1, P_2, P_3$) connecting a source and a target node. Each path, represented as a sequence of nodes, is passed through shared embedding layers followed by stacked BiLSTM layers. The resulting features are aggregated using global max pooling and a dense fusion layer to form a unified path representation. The final output is the sigmoid probability of the target relationship $R$, computed as the dot product of the normalized path and edge embeddings.}
    \label{fig:nn}
\end{figure}

To ensure the inductive nature of the method, we maintain strict isolation between the training and testing environments. Specifically, the model is trained on a graph generated from one set of execution scenarios and evaluated on graphs derived from entirely different scenarios. While these graphs share a common set of relationships (predicates), their node sets are mutually exclusive. 
To enable generalization to unseen entities, we represent each path as a sequence of edge types rather than specific node identifiers. This focus on edge sequences is central to our inductive approach, allowing the neural network (NN) to learn structural patterns that remain valid even when encountering entirely new nodes in the database environment. The NN architecture is presented in Fig.~\ref{fig:nn}.

%========================================================
% We explore six different modern approaches to Inductive Link Prediction (ILP):
% \begin{enumerate}
%     \item LSTM path embedding
%     \item Variational Graph Auto-encoders
%     \item Learnable Topology Augmentation
%     \item Knowledge Distillation
%     \item Generative Interpretation Networks
%     \item Graph Attention Networks
% \end{enumerate}

% This choice is dictated by the performance in the general ILP tasks.

\subsection{Results}
%-----------------------------------
The evaluation is conducted on $9$ different tasks, covering various types of transformations (no transformation, linear, and non-linear) as well as algebraic operations (selection, join, and union). Each task comprises $20$ scenarios, of which $17$ are used to construct the training data. For this purpose, we prepare a database containing $17$ transformation scenarios and convert it into a KG. For each edge (relationship) in the KG, three random paths are sampled, resulting in a large number of training samples (approximately $5$ million per graph), which are used to train the NN. The training is performed for three epochs with a batch size of $32$.
The remaining 3 scenarios are used for testing. In the test graph, the \textit{data lineage} relationships are removed and used only as ground truth. For evaluation, we sample $4000$ negative examples consisting of pairs of nodes corresponding to database rows, ensuring that no \textit{rowDerivedFrom} relationship exists between them. The positive samples include all node pairs connected by the \textit{rowDerivedFrom} relationship. The NN is then tasked with distinguishing between positive and negative samples.

\begin{table}[]
    \centering
    \caption{Experimental comparison of data lineage prediction performance based on KGs constructed using the baseline and RDDL ontologies. }
\begin{tabular}{lcllll}
      \hline
     \textbf{Task} & \textbf{Ontology} & \textbf{Precision} & \textbf{Recall} & \textbf{AUC} & \textbf{Hits@10} \\
     \hline
     Selection-projection & baseline & 0.92 & 0.95 & 0.98 & 0.62 \\
     Selection-projection & RDDL & 0.95(+0.03) & 0.95(-) & 0.98(-) & 0.71(+0.09) \\
     Selection-linear & baseline &  0.94 & 0.89 & 0.98 & 0.59 \\
     Selection-linear & RDDL & 0.96(+0.01) & 0.96(+0.07) & 0.98(-) & 0.74(+0.15) \\
     Selection-nonlinear & baseline & 0.91 & 0.89 & 0.97 & 0.68 \\
     Selection-nonlinear & RDDL & 0.95(+0.04) & 0.90(+0.01) & 0.98(+0.01) & 0.72(+0.04) \\
     Join-projection & baseline &  0.82 & 0.84 & 0.83 & 0.45 \\
     Join-projection & RDDL &   0.88(+0.06) & 0.91(+0.07) & 0.89(+0.06) & 0.68(+0.23) \\
     Join-linear & baseline &   0.79 & 0.81 & 0.78 & 0.42 \\
     Join-linear & RDDL &   0.83(+0.04) & 0.81(-) & 0.84(+0.06) & 0.62(+0.20) \\
     Join-nonlinear & baseline &   0.85 & 0.83 & 0.83 & 0.31 \\
     Join-nonlinear & RDDL &  0.84(-0.01) & 0.85(+0.02) & 0.86(+0.03) & 0.57(+0.26) \\
     Union-projection & baseline &  0.86 & 0.87 & 0.88 & 0.68 \\
     Union-projection & RDDL &  0.94(+0.08) & 0.91(+0.04) & 0.91(+0.03) & 0.71(+0.03) \\
     Union-linear & baseline &  0.88 & 0.89 & 0.90 & 0.62 \\
     Union-linear & RDDL &  0.95(+0.07) & 0.93(+0.04) & 0.95(+0.05) & 0.67(+0.05) \\
     Union-nonlinear & baseline &  0.87 & 0.91 & 0.90 & 0.67 \\
     Union-nonlinear & RDDL &  0.93(+0.06) & 0.91(-) & 0.91 (+0.01) & 0.66 (-0.01) \\
     \hline
     \multicolumn{2}{l}{\textbf{Average improvement}}  &  \textbf{\ \ \ 0.04} & \textbf{\ \ \ 0.02} & \textbf{\ \ \ 0.03} & \textbf{\ \ \ 0.11} \\
     \hline
\end{tabular}
    \label{tab:results}
\end{table}

In Table~\ref{tab:results}, we report the test results using standard machine learning metrics, including \textit{Precision}, \textit{Recall}, \textit{Area Under the Precision--Recall Curve} (AUC), and \textit{Hits@10}. Due to the high density of the graph, comparing each edge from the set of positive samples against all possible node pairs is computationally prohibitive. Consequently, we adopt a ranking-based evaluation, where each positive sample is ranked against a pool of negative samples. We report \textit{Hits@10}, defined as the frequency with which the ground-truth positive sample appears within the top ten positions of the ranking.

For each task, we report the performance of both approaches, while the improvements achieved by RDDL over the baseline are shown in parentheses next to the corresponding metric values.
Overall, the RDDL ontology consistently improves performance across most tasks and metrics, with particularly strong gains observed for \textit{Hits@10}. The final row summarizes the results as the average improvement across all tasks, computed as a simple mean of the per-task differences. %These results confirm that the proposed ontology provides a more effective semantic representation for knowledge graph-based data lineage prediction.

\section{Conclusion and Future Work}
%========================================================
In this paper, we proposed \textit{RDDL Ontology} - an extended ontology for \textit{data lineage} in relational databases, designed to provide a more expressive and semantically rich representation than the model proposed in our previous work \cite{dutkiewicz2026}. The new ontology incorporates additional concepts, including a refined class hierarchy, execution semantics as well as detailed modeling of constraints and data types. We evaluated its impact on data lineage discovery using a KG-based inductive link prediction framework. We observed that employing a semantically richer model enriched the KG's structure, increasing its density and expanding the diversity of both relationship types (edges) and modeled classes. We confirmed that the use of this enhanced graph enables higher precision in discovering broken data lineage links. The experimental results demonstrate that the application of the proposed ontology improves the performance of lineage link prediction, with a mean gain of \textbf{$3\%$} in terms of \textit{AUC} and \textbf{$11\%$} in terms of \textit{Hits@10} over the baseline.
% , and that combining it with a more advanced Graph Attention Network further enhances predictive accuracy.

In the nearest future, we first intend to perform a granular analysis of how individual ontology components influence link prediction performance, specifically through ablation studies across diverse lineage scenarios. Furthermore, we will further refine the ontology and evaluate its applicability in more complex scenarios with the use of broader range of inductive link prediction models. Finally, we will enhance our experimental evaluation protocol to ensure it more closely mirrors the challenges of real-world missing lineage discovery in relational databases.

\bibliographystyle{splncs04}
\bibliography{biblio}
\end{document}